\documentclass[letterpaper,journal]{IEEEtran}

\usepackage{amsmath,amsfonts}

\usepackage{amsmath} 
\usepackage[linesnumbered,ruled,vlined]{algorithm2e}
\usepackage{array}
\usepackage[caption=false,font=normalsize,labelfont=sf,textfont=sf]{subfig}
\usepackage{textcomp}
\usepackage{stfloats}
\usepackage{url}
\usepackage{verbatim}
\usepackage{graphicx}
\usepackage{cite}
\usepackage{amssymb}
\graphicspath{ {./fig/} }
\usepackage{xcolor}
\begin{document}

\title{Implicit Semantic Communication Based on Bayesian Reconstruction Framework}

\author{Yiwei Liao, Shurui Tu, Yujie Zhou, Dongzi Jin, Yong Xiao, Yingyu Li

\thanks{*This work is accepted at IEEE Wireless Communications Letters. Copyright
may be transferred without notice, after which this version may no longer
be accessible.

This work was supported in part by the Mobile Information Network National Science and Technology Key Project under grant 2024ZD1300700, and the National Natural Science Foundation of China (NSFC) under grants 62525109 and 62301516. (Corresponding author: Yong Xiao.)

Yiwei Liao, Shurui Tu, Yujie Zhou, Dongzi Jin, and Yong Xiao are with the School of Electronic Information and Communications, Huazhong University of Science and Technology, Wuhan 430074, China. Yong Xiao is also affiliated with Peng Cheng Laboratory, Shenzhen, China, and Pazhou Laboratory (Huangpu), Guangzhou, China (e-mail: \{liao\_yiwei, shurui\_tu, zhouyujie2357, jdzsuper, yongxiao\}@hust.edu.cn). Yingyu Li is with the School of Mechanical Engineering and Electronic Information, China University of Geosciences, Wuhan, China (e-mail: liyingyu29@cug.edu.cn). \\
Code available at: \texttt{https://github.com/Yiwei-Liao/SBRF}
}
}

\maketitle

\begin{abstract}
Semantic communication is a novel communication paradigm that focuses on the transportation and delivery of the \emph{meaning} of messages. 
Recent results have verified that a graphical structure provides the most expressive and structurally faithful formalism for representing the relational semantics in most information sources. However, most existing works represent the semantics based on pairwise relation-based graphs, which cannot capture the higher-order interactions that are essential for some semantic sources. This paper proposes a novel Bayesian hypergraph inference-based semantic communication framework that can directly recover implicit semantic information involving high-order hyperedges at the receiver based on the pairwise relation-based explicit semantics sent by the transmitter. 
Experimental results based on real-world datasets demonstrated that the proposed SBRF achieves up to 90\% recovery accuracy of the high-order hyperedges based on the pairwise relation-based explicit semantics. 
\end{abstract}

\begin{IEEEkeywords}
Semantic communication, implicit semantics, hypergraph, inference, Bayesian estimation
\end{IEEEkeywords}

\section{Introduction}

Semantic communication, a novel communication paradigm that focuses on transmitting the essential \emph{meaning} of the source signal instead of its bit-level representations, has recently emerged as a promising solution for the next generation of communication systems~\cite{yang2022semantic,shi2024introduction}. Previous works have already verified that modeling the semantic knowledge of the source signal as a graphical structure can provide the most expressive and structurally faithful formalism of the relational semantics inherent in any information source~\cite{xiao2023reasoning, seo2023semantics}. However, most existing works adopt the pairwise relation-based graphical structure to represent the semantic information, while ignoring the higher-order, multi-entity interactions that are essential for the understanding and interpretation of semantic information sources~\cite{battiston2020networks}.


Recent studies have indicated that the majority of real-world networking systems inherently exhibit high-order interactions among their constituent knowledge entities. These higher-order semantic relations are more suitable to be represented as hypergraph structures~\cite{chen2023multi,antelmi2023survey}. Unfortunately, acquiring or directly observing these crucial high-order interactions from extant information sources presents significant methodological challenges. There is a critical need to develop simple and effective solutions capable of inferring and accurately reconstructing the latent high-order interactions inherent in these systems, utilizing only the limited empirical data derived from readily available pairwise relational observations. 

Motivated by the above observation, in this paper, we propose the Semantic Bayesian Reconstruction Framework (SBRF), a novel semantic communication architecture that employs Bayesian hypergraph inference to recover implicit semantic information from observed pairwise relations. In SBRF, the semantic encoder can only recognize explicit semantics, which consist of the key semantic knowledge entities and the pairwise relations between these entities. It will focus on compressing these explicit semantics for physical channel transmission. The semantic decoder not only recovers the explicit semantic sent by the semantic encoder, but also learns an inference mechanism that can infer the implicit semantic information, involving high-order interactions and hyperedges that are critical for understanding the real semantic intention based on the recovered explicit semantics and some prior knowledge about the signal source. We propose a computationally efficient algorithm that iteratively refines the hyperedge with the highest likelihood. We prove that the computational complexity of our proposed algorithm scales linearly with the maximum size of the recoverable hyperedge. Extensive experiments have been conducted based on real-world datasets. Our results show that the proposed SBRF achieves up to 90\% recovery accuracy of the high-order hyperedges based on the received explicit semantics. 

\section{System Model and Problem Formulation}

We follow a commonly adopted setting\cite{xiao2023reasoning} and represent the semantics of a given information source as a triple: $\langle \mathcal{G}, \mathcal{H}, \Phi \rangle$, where $\mathcal{G}$ corresponds to the {\em explicit semantic information}, $\mathcal{H}$ is the {\em implicit semantic information}, and $\Phi$ is the inference mechanism that maps the explicit semantics into the implicit semantic information. In this paper, we consider a graph-based representation of semantics in which $\mathcal{G}$ corresponds to a set, involving the semantic knowledge entities and their relationships that are directly observable from the source signal. $\mathcal{H}$ includes the implicit connection and correlations among entities that are critical for the understanding of semantic information, but cannot be directly recognized from the source. Motivated by the fact that most real-world networking systems, although they exhibit high-order interactions among constituent entities, are often represented by pairwise relations, in this paper, we model the explicit semantics as a normal graph that consists of only the directly observable entities and pairwise relations, and the implicit semantics as hypergraphs, involving higher-order interactions that can be inferred from the explicit semantics. More formally, let $\mathcal{G}= \langle \mathcal{V},\mathcal{R} \rangle$, where $\mathcal{V}=\{v_1,v_2,\ldots,v_{|\mathcal{V}|}\}$ is the set of observable entities and $\mathcal{R}=\{r_{ij}=(v_i,v_j)\mid v_i,v_j\in\mathcal{V}\}$ is the set of pairwise relations. Also, let $\mathcal{H}=\langle \mathcal{V},\mathcal{E} \rangle$ be the implicit semantic, where $\mathcal{E}$ is the set of the high-order semantic relations $e$ that can be inferred from $\mathcal{G}$.  
Let $\Phi$ be an inference mechanism that generates $\mathcal{H}$ based on $\mathcal{G}$.    


In this paper, we consider a generative model-based semantic communication, in which the transmitter focuses on recognizing and compressing the explicit semantics for efficient physical channel transmission, and the main objective of the receiver is to recover the implicit semantics based on the received explicit semantics and some prior information about the information source, as illustrated in Fig. \ref{fig:system_model}. 


\section{Semantic Bayesian Reconstruction Framework} 

\begin{figure*}[htbp]
    \centering
    \includegraphics[width=1\textwidth]{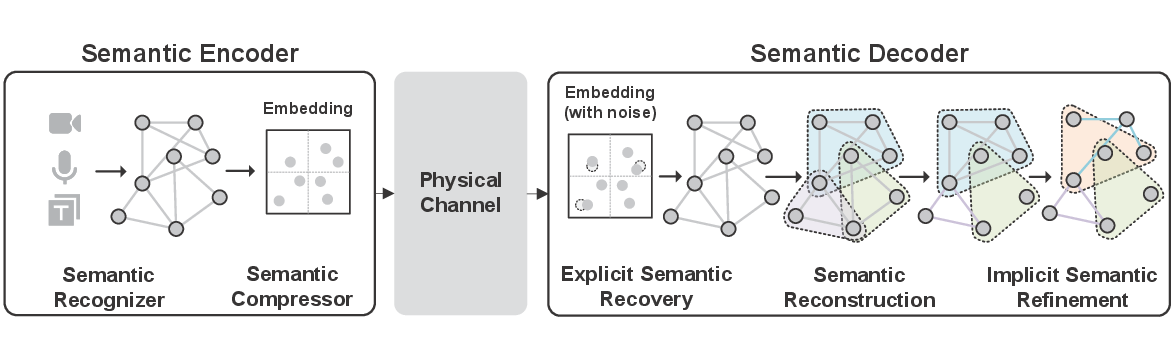}
    \caption{The framework of our proposed semantic communication architecture.}
    \label{fig:system_model}
\end{figure*}

We propose the Semantic Bayesian Reconstruction Framework (SBRF), a semantic communication architecture that employs Bayesian hypergraph inference to recover implicit semantic information from the observed pairwise relation-based graphs. The overall framework is illustrated in Fig.~\ref{fig:system_model} and the major components are described in detail as follows: 

\subsection{Semantic Encoder}

This module consists of the following components.  

\subsubsection{Semantic Recognizer}

The semantic recognizer identifies the key knowledge entities and relations from the source signal and converts the identified entities and relations into a graph $\mathcal{G}=(\mathcal{V},\mathcal{R})$, where $\mathcal{V}$ represents the set of semantic entities, and the edges in $\mathcal{R}$ represent the pairwise interactions between any pair of entities. 




\subsubsection{Semantic Compressor}
To efficiently transmit the recognized graph into the physical channel, the semantic compressor converts the graph into a low-dimensional representation space, called the embedding space, and then encodes the low-dimensional embeddings into suitable forms to be sent to the channel. For example, in our previous work\cite{xiao2023reasoning}, we have developed a projection-based encoding function to map the explicit semantics into the semantic constellation space, in which the low-dimensional embeddings can be sent either via a sequence of real-valued signals with
$l_1$ norm distance metric, e.g., using the amplitude modulation scheme, or complex-valued signals with $l_2$ norm distance metric using phase and amplitude modulation schemes. 

\subsection{Semantic Decoder}

The semantic decoder is composed of the following key components. 


\subsubsection{Explicit Semantic Recovery}

The semantic decoder first obtains a recovered explicit semantic information, denoted as $\hat{\mathcal{G}}$, based on the received signal. Note that the embeddings received by the decoder can be corrupted during the physical channel transmission. These corrupted embeddings can be recovered by using the pairwise-relation-based inference methods. In this case, some redundancy is required when compressing explicit semantics into low-dimensional embeddings in the semantic compressor of the encoder, as described in \cite{xiao2023reasoning}.


\subsubsection{Hypergraph-based Implicit Semantic Reconstruction} 

Based on the recovered pairwise-relational graph $\hat{\mathcal{G}}$, we propose a Bayesian generative model-based solution to infer the underlying hypergraph structures that are hidden from the source signal, but are essential for understanding the real semantic intention of the source user. 
Formally, we follow a commonly adopted setting~\cite{xia2022bayesian} and assume that some prior information about the source user is available. This prior information may correspond to semantic constraints or relational patterns derived from existing knowledge bases or other available background information, such as ontologies or domain-specific knowledge base. Let $P(\mathcal{H}|\theta)$ be the empirical prior probability of hypergraph $\mathcal{H}$, where $\theta$ represents a set of latent variables or parameters associated with the semantic information sources. 

Then we can calculate the posterior probability of hypergraph $\mathcal{H}$ based on the observed graph $\hat{\mathcal{G}}$ and the prior information as follows:
\begin{equation}
    P(\mathcal{H}|\hat{\mathcal{G}},\theta) = \frac{P(\hat{\mathcal{G}}|\mathcal{H},\theta)P(\mathcal{H}|\theta)}{\sum_{\mathcal{H'}}P(\hat{\mathcal{G}}|\mathcal{H'},\theta)P(\mathcal{H'}|\theta)}
    \label{eq_posteriorH}
\end{equation}
where the likelihood term is given by:
\begin{equation}
P(\hat{\mathcal{G}}|\mathcal{H},\theta)=\prod_{(i,j)\in\hat{\mathcal{E}}'}p^{A_{ij}}(1-p)^{1-A_{ij}}
\end{equation}
with $A_{ij}$ is the observed adjacency (1 if edge exists, 0 otherwise), and $p$ is the conditional probability of observing an edge given a hypergraph structure $\mathcal{H}$. 

In case that only a subset of entities ${\cal V}_{\Omega}$ can be recovered by the explicit semantic recovery components, we can use the following likelihood function to retrieve the posterior probability of the hypergraph $\mathcal{H}$: 

\begin{footnotesize}

\begin{eqnarray}
P(\hat{\cal G}_{\Omega}\mid \mathcal{H},\theta)=\prod_{(v_i, v_j)\in {\cal V}_{\Omega}} \left[1-(1-p)^{|\mathcal{E}_{ij}|}\right]^{A_{ij}}\left[(1-p)^{|\mathcal{E}_{ij}|}\right]^{1-A_{ij}}
\label{eq:partial_likelihood}
\end{eqnarray}

\end{footnotesize}

where ${\hat{\cal G}}_{\Omega}$ is a subset of $\hat{\cal G}$ consisting of entities in ${\cal{V}}_{\Omega}$ and the corresponding edges, $p$ is the probability that an individual hyperedge generates a pairwise interaction, and $\mathcal{E}_{ij}$ denotes the set of hyperedges in $\mathcal{H}$ containing both entities $i$ and $j$. $|\mathcal{E}_{ij}|$ is the number of hyperedges in $\mathcal{E}_{ij}$. To simplify our notation, in the rest of this paper, we focus on the case that $\hat{\mathcal{G}}$ can be recovered by the explicit semantic recovery component.

We can observe that, even when only partial structural information is available at the transmitter, our method directly recovers the implicit high-order hypergraphs at the receiver through Bayesian inference.


\subsubsection{Bayesian-inference-based Implicit Semantic Refinement} 


We can observe that the hypergraph $\cal H$ generated from the posterior calculated in equation (\ref{eq_posteriorH}) may consist of redundant or low possible hyperedges, e.g., the same hyperedges can be generated repeatedly from pairwise edges observed in the explicit semantics. Therefore, in this subsection, 
we introduce a Bayesian-inference-based implicit semantic refinement component that removes the redundant hyperedges and keeps the most probable hyperedges according to the implicit semantic constraints, i.e., our main objective is to obtain the optimal candidate of the implicit semantics $\mathcal{H}^*$ as follows:

\begin{equation}
\mathcal{H}^*=\arg\max_{\mathcal{H}}P(\mathcal{H}|\hat{\mathcal{G}},\theta).
\label{eq_OptimalH}
\end{equation}

Specifically, we adopt a Markov Chain Monte Carlo (MCMC) method to iteratively sample candidate hypergraphs by probabilistically adding, removing, or modifying hyperedges. This iterative updating enables convergence toward the most probable hypergraph configuration, maximizing the posterior probability $P(\mathcal{H}|\hat{\mathcal{G}},\theta)$.


To capture the relationships between observed pairwise interactions and latent hyperedges, we define the likelihood function $P(\hat{\cal G}|\mathcal{H},\theta)$, assuming that any two hyperedges are conditionally independent under the given $\hat{\cal G}$. 

Suppose the likelihood of observing each pairwise edge $(i,j)$ in the explicit semantics $\mathcal{G}$ based on the prior information of $\mathcal{H}$ is given by,  
\begin{equation}
P(A_{ij}=1|\mathcal{H},\theta) = 1-(1-p)^{|\mathcal{E}_{ij}|}
\end{equation}

Then, the likelihood of $\hat{\mathcal{G}}$ under given $\mathcal{H}$ can be written as: 

\begin{equation}
P(\hat{\mathcal{G}}|\mathcal{H},\theta)=\prod_{r_{ij} \in\mathcal{R}}[1-(1-p)^{|\mathcal{E}_{ij}|}]\prod_{r_{ij} \notin\mathcal{R}}(1-p)^{|\mathcal{E}_{ij}|}
\end{equation}

As mentioned earlier, the same hyperedges can be generated repeatedly from some pairwise edges, i.e., let $\delta(e)$ be the number of the same hyperedges generated by the explicit semantics and $\delta(e) \ge 1$. We then introduce a hypergraph posterior distribution $P(\mathcal{H}|\hat{\mathcal{G}},\theta)$ that penalizes structural redundancy, favoring the hypergraphs with higher possibilities, e.g., we follow the same line as \cite{battiston2020networks} and incorporate the hyperedge multiplicity and size constraints by applying the following posterior distribution:

\begin{small}
\begin{equation}
P(\mathcal{H}|\hat{\mathcal{G}},\theta) \propto \exp\left[-\beta|\mathcal{E}|-\gamma\sum_{e\in\mathcal{E}}(\delta(e)-1)\right]\mathbb{I}(|e|\leq L)
\label{eq_LengthConstraint}
\end{equation}
\end{small}

where $|\mathcal{E}|$ is the total number of hyperedges, 
parameters $\beta$ and $\gamma$ enforce sparsity and uniqueness, respectively, and the indicator $\mathbb{I}(|e|\leq L)$ constrains the size of hyperedges.

To calculate the optimal candidate of the implicit semantics that maximizes the hypergraph posterior distribution in (\ref{eq_OptimalH}), we utilize the Metropolis-Hastings (MH) algorithm to iteratively sample the candidate hypergraphs based on the random walk. More specifically, in each iteration $t$, a candidate hypergraph $\mathcal{H}'$ is proposed by adding, deleting, or modifying a hyperedge based on the accepting probability $\alpha$, calculated by

\begin{equation}
\hspace{-4mm}
\alpha = \min\left[1,\frac{P(\hat{\mathcal{G}}|\mathcal{H}^{(t)},\theta)P(\mathcal{H}^{(t)}|\theta)Q(\mathcal{H}^{(t-1)}|\mathcal{H}^{(t)})}{P(\hat{\mathcal{G}}|\mathcal{H}^{(t-1)},\theta)P(\mathcal{H}^{(t-1)}|\theta)Q(\mathcal{H}^{(t)}|\mathcal{H}^{(t-1)})}\right]
\end{equation}
where $Q(\mathcal{H}^{(t-1)}|\mathcal{H}^{(t)})$ is the probability distribution of proposing the move from $\mathcal{H}^{(t-1)}$ to $\mathcal{H}^{(t)}$.

The above algorithm is illustrated in Algorithm \ref{alg:mcmc}. 
\begin{algorithm}[htbp]
\caption{MCMC Hypergraph Reconstruction}
\label{alg:mcmc}
%
Initialize hypergraph \(\mathcal{H}^{(t=0)}\) generated from reconstruction\;
\For{\(t = 1, 2, \dots, T\)}{
    Randomly select a hyperedge candidate \(e\)\;
    Randomly add or remove a sub-hyperedge of \(e\)\;
    Propose new hypergraph \(\mathcal{H}^*\) by applying the change\;
    Compute acceptance ratio $\alpha$;\\
    Generate random number \(u \sim \text{Uniform}(0,1)\)\;
    \If{\(u < \alpha\) and \(P(\hat{\cal G} |\mathcal{H}^*)=1\)}{
        Set \(\mathcal{H}^{(t)} \leftarrow \mathcal{H}^*\)\;
    }
    \Else{
        Set \(\mathcal{H}^{(t)} \leftarrow \mathcal{H}^{(t-1)}\)\;
    }
}
\Return{posterior samples \(\{\mathcal{H}^{(t)}\}_{t=1}^{T}\)\;}
\end{algorithm}

\subsection{Complexity Analysis}
To evaluate the computational complexity of our proposed Bayesian inference framework, we analyze the cost to perform each MCMC iteration. More specifically, in each iteration, the algorithm proposes a new hypergraph structure by adding, removing, or modifying a hyperedge with a maximum size constraint $L$ according to (\ref{eq_LengthConstraint}). When proposing a modification to a hyperedge $e$ involving $k$ entities for $k \le L$, the update of the posterior probability (as defined in previous equations) only requires to calculate the subset of entities within this hyperedge, 
resulting in a computational complexity of $O(k^2)$. 
This complexity is bounded by $O(L^2)$ per iteration, which is independent of the total number of entities or hyperedges in the hypergraph.

In practice, considering the hyperedge size $L$ is generally small and fixed by design, our proposed algorithm is computationally efficient and scalable to large graphs, making our proposed approach practically suitable for real-time semantic communication systems with complex semantic information sources. Since each iteration of the MCMC sampler updates a single hyperedge, the overall convergence time of our Bayesian inference procedure scales approximately linearly with the total number of hyperedges $|\mathcal{E}|$, a trend also confirmed by our empirical convergence analysis.
Therefore, the overall complexity of our algorithm is given by $O(|\mathcal{E}|L^2)$, ensuring feasibility for large-scale hypergraphs. 

\section{Experimental Results}


We conduct extensive experiments based on six real-world datasets: FB-AUTO, JF17K, M-FB15k, Wikipeople, NDC\_C, and Walmart, to validate the effectiveness of our proposed SBRF. 

\begin{figure}
  \begin{minipage}{0.48\linewidth}
  \vspace{0.1in}
   \centering
   \includegraphics[width=\linewidth]{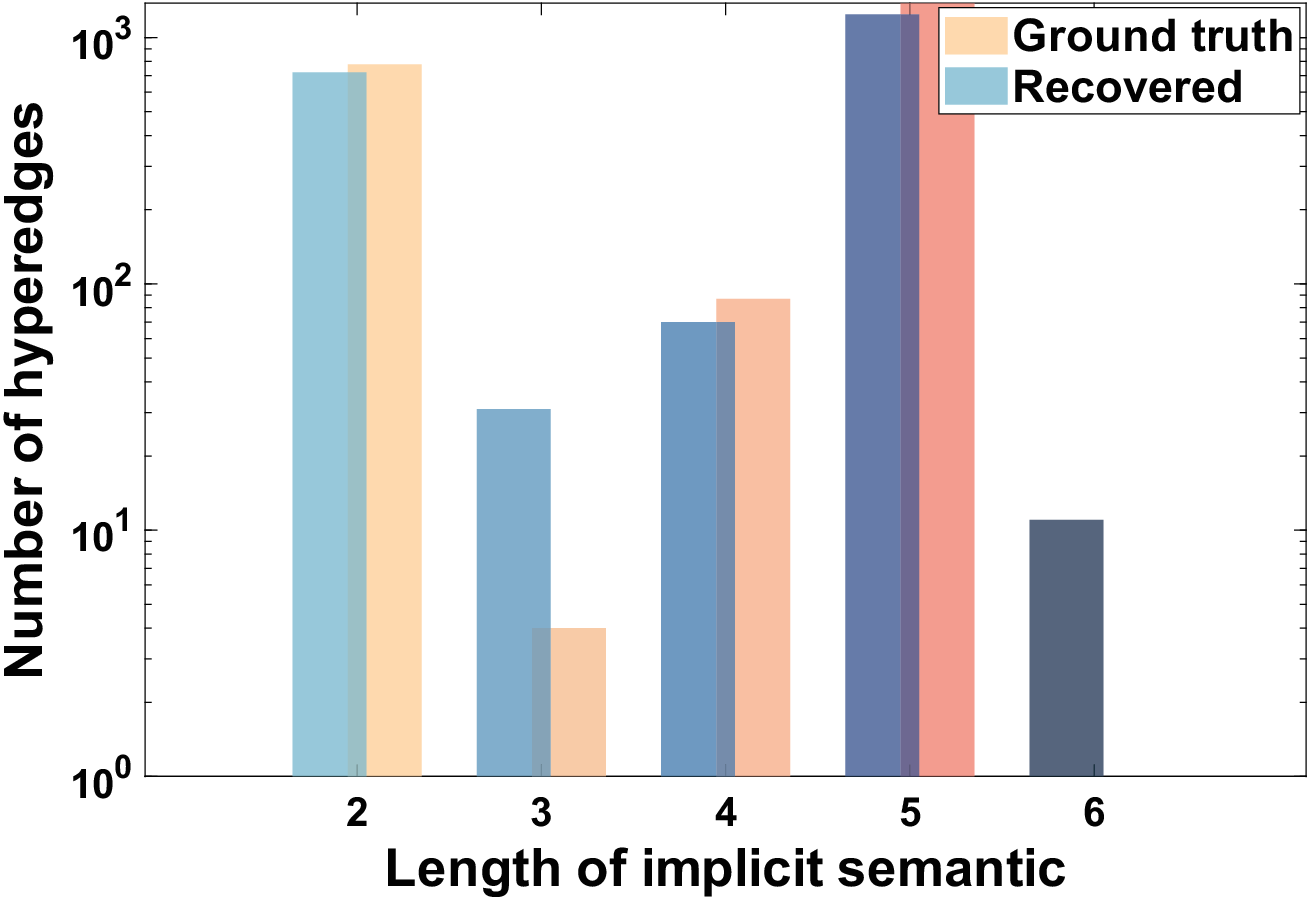}
  \caption{The number of ground truth and recovered hyperedges with different numbers of entities based on the experiments conducted on the FB-AUTO dataset.}
  \label{lenth_volume}
  \end{minipage}
  \begin{minipage}{0.48\linewidth}
   \centering
   \includegraphics[width=\linewidth]{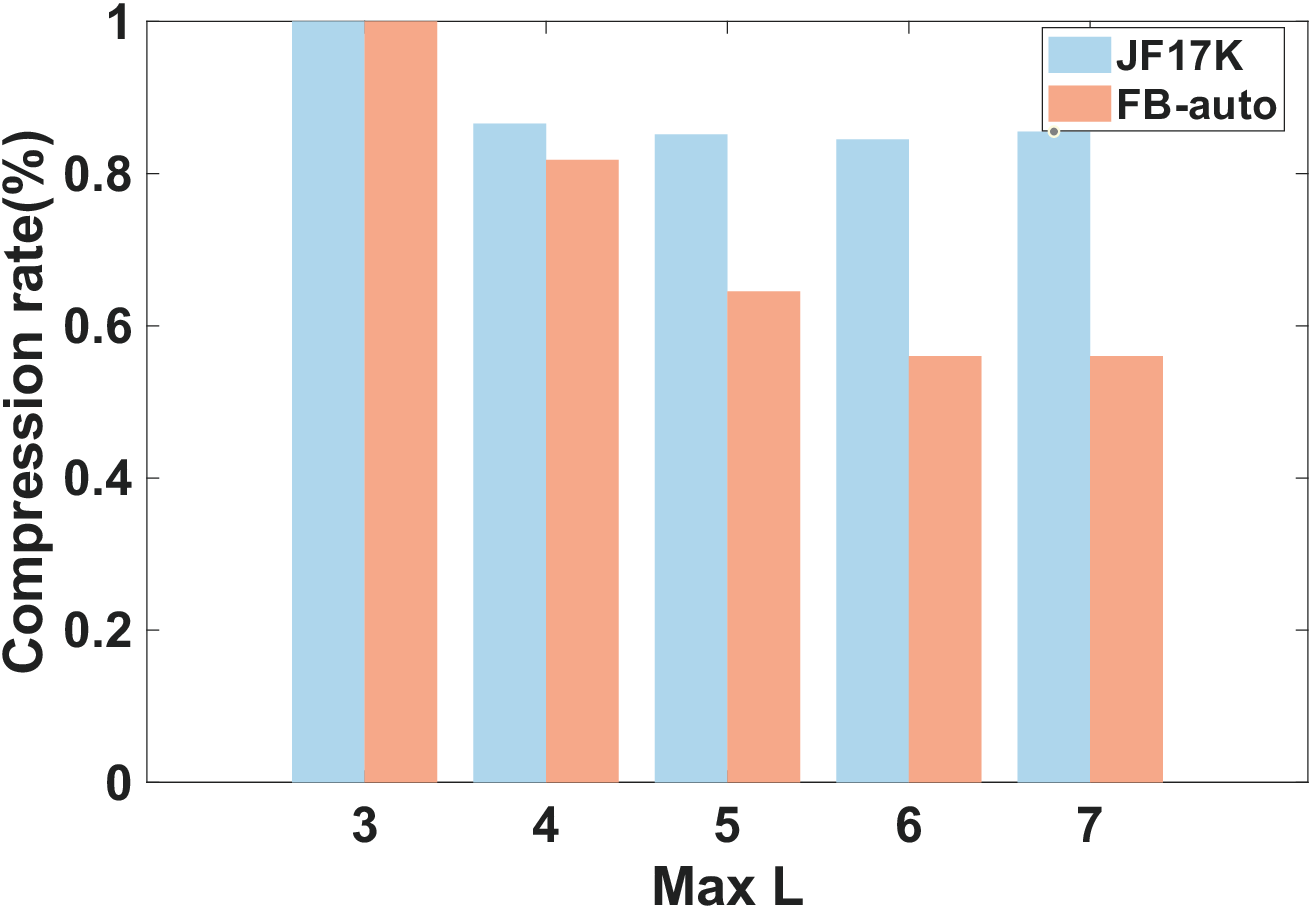}
  \caption{The compression rate of the semantic compressor at the semantic encoder under different limits of maximum hyperedge lengths $L$. } 
  \label{lenth_limit}
  \end{minipage}

\end{figure}

We first demonstrate the effectiveness of SBRF in recovering implicit semantic relationships with different lengths, i.e., the number of entities in each hyperedge, based on the received embeddings. In Fig.~\ref{lenth_volume}, we compare the number of ground truth and recovered hyperedges with different numbers of entities based on the experiments conducted on the FB-AUTO dataset. We can observe that our proposed Bayesian inference method can successfully reconstruct the hyperedges with different lengths, effectively recovering the essential implicit semantic information at the receiver.


In Fig.~\ref{lenth_limit}, we present the compression rate, i.e., the ratio of the file size of the original hypergraph of the information source and that of the low-dimensional embeddings calculated based on our proposed semantic compressor, of the semantic compressor at the semantic encoder under different limits of maximum hyperedge lengths $L$ based on the experiments conducted by two datasets JF17K and FB-AUTO. We can observe that, for both datasets, the compression rate is low, i.e., approaches 1, when the maximum hyperedge lengths are restricted to a smaller value, i.e., $L=2, 3$. This is because when the recovered hyperedge lengths are limited to 2 (regular pairwise edge) or 3, the recovered semantics will be dominated by the explicit semantic information, and the higher-order implicit semantics, especially those with higher-order relations (hyperedges with lengths exceeding 4 or higher values), cannot be recovered. Also, because the FB-AUTO dataset has less number of entities but more high-order hyperedge ($L\geq 6$) than JF17K, the compression rate of FB-AUTO is always higher than JF17K for a given constraint on the recovered hyperedge sizes. We can also observe that when the limits of the recovered hyperedge sizes increase, the compression rates of dataset JF17K increase at first and then decrease when hyperedge sizes exceed 6. This means that to achieve the maximum compression rate of a given dataset, it is important to find the appropriate constraints on the recovered hyperedge size limits.





\begin{figure}
  \begin{minipage}{0.49\linewidth}
   \centering
   \includegraphics[width=\linewidth]{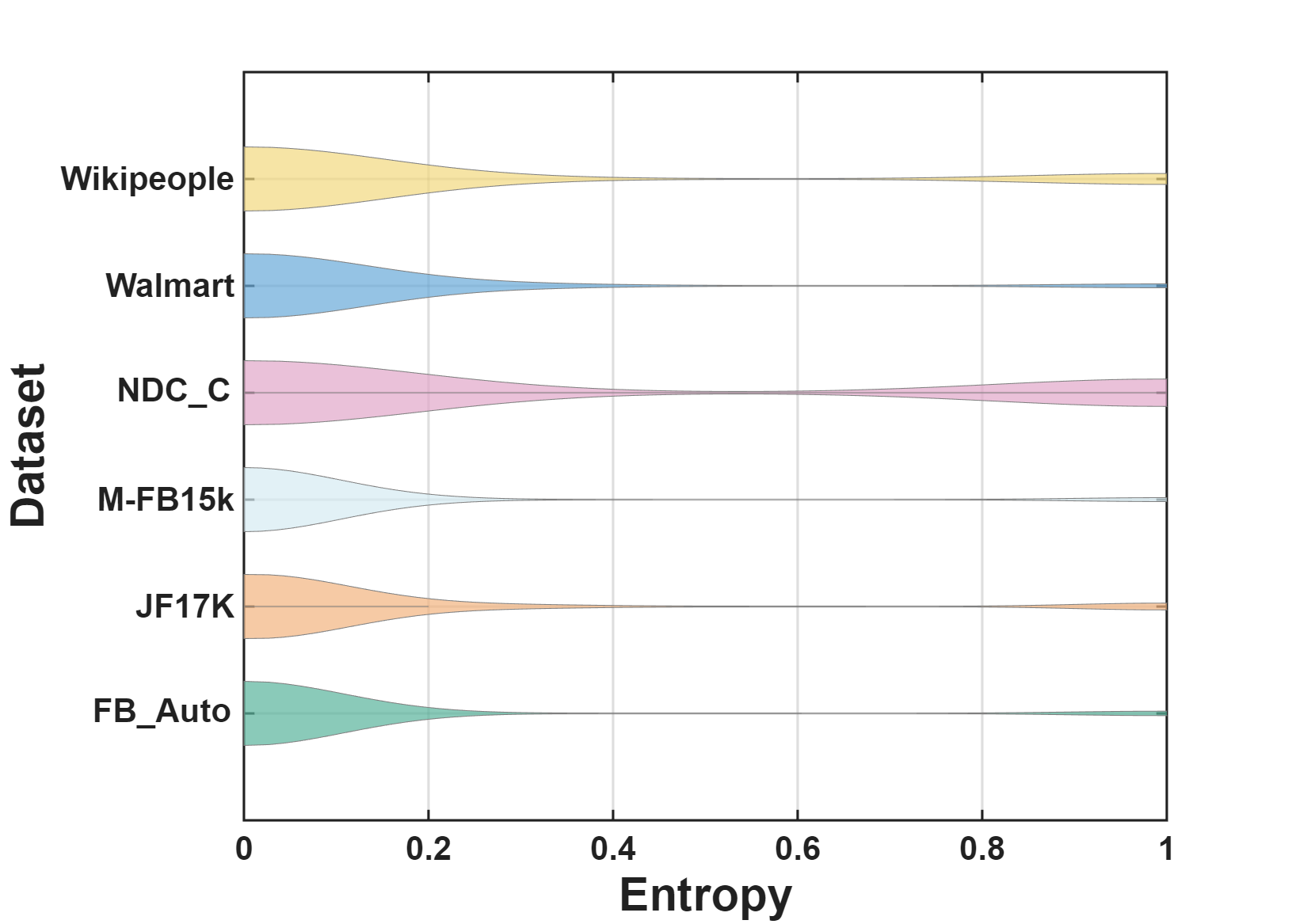}
  \caption{The entropy values of the accepting probability of our proposed MH algorithm for six different datasets: FB-AUTO, JF17K, M-FB15k, Wikipeople, NDC\_C, and Walmart. } 
  \label{entropy}
  \end{minipage}
  \begin{minipage}{0.48\linewidth}
   \centering
   \includegraphics[width=\linewidth]{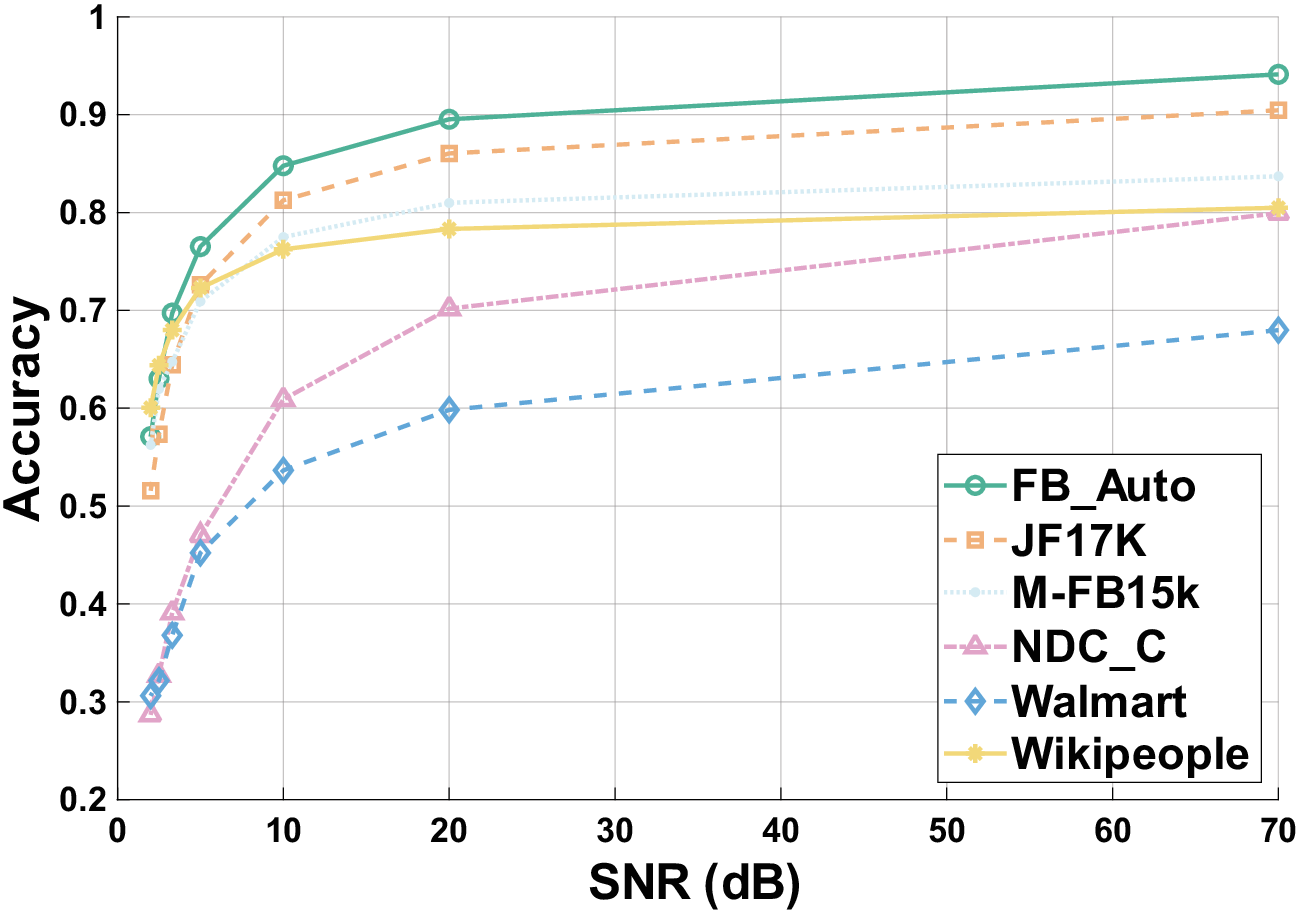}
  \caption{The accuracy of implicit semantic information recovery under different received SNRs based on the experiments conducted on six different datasets. } 
  \label{noise}
  \end{minipage}
\end{figure}

In Fig.~\ref{entropy}, we present the violin plot showing the distribution of entropy values of the accepting probability in our proposed MH algorithm for six different datasets, including FB-AUTO, JF17K, M-FB15k, Wikipeople, NDC\_C, and Walmart. The wider sections represent a higher frequency of accepting probabilities at that entropy level. We can observe that the entropy values of all six datasets are concentrated in either the near 1 (accept) or the near 0 (reject) regions, indicating that in each iteration of the random walk, each proposed hyperedge refinement is either accepted or rejected with high confidence.


In Fig.~\ref{noise}, we present the accuracy of implicit semantic information recovery under different received signal-to-noise ratios (SNRs) from $-10$ dB to $70$ dB based on the experiments conducted on six different datasets. We can observe that with the increase of the SNR, the recovery accuracies of our proposed methods increase in all six datasets. Also, the recovery accuracy reaches up to 90\% even when the SNR is as low as 20 dB. This justifies the capability for accurately recovering implicit semantics with high-order relations of our proposed SBRF even at very low SNR scenarios.


\begin{figure}
    \centering
    \includegraphics[width=1\linewidth]{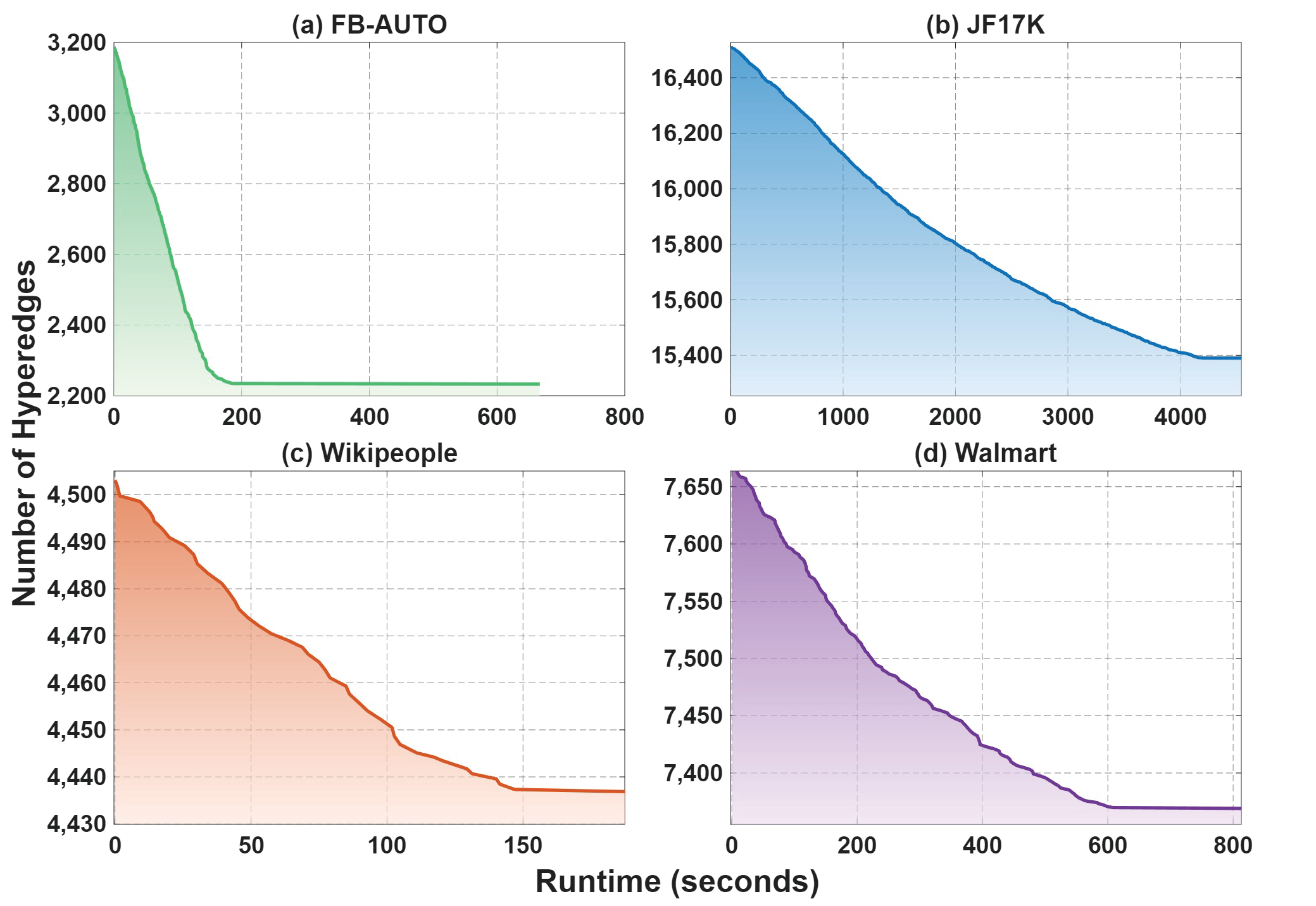}
    \caption{The number of hyperedges left in our MH algorithm under different runtimes based on experiments conducted at four datasets: (a) FB-AUTO, (b) JF17K, (c) Wikipeople, and (d) Walmart.}
    \label{fig:convergence_runtime}
\end{figure}

In Fig.~\ref{fig:convergence_runtime}, we present the numbers of hyperedges left in our MH algorithm under different runtimes, indicating different refinement speeds, i.e., the number of hyperedges that have been removed under different runtimes, based on experiments conducted at four datasets: (a) FB-AUTO, (b) JF17K, (c) Wikipeople, and (d) Walmart. We can observe that in all four datasets, the number of hyperedges that are left at the end of the iterations approaches a static value in all the different datasets. This means that our proposed algorithm can always converge to the optimal set of hyperedges with the highest probabilities. We can always observe that the refinement speeds of our algorithm are almost linear with the runtime. This verifies our observation in the complexity analysis, which shows that the computational complexity of our algorithm increases linearly with the total number of hyperedges. 


\begin{figure}
    \centering
    \includegraphics[width=1\linewidth]{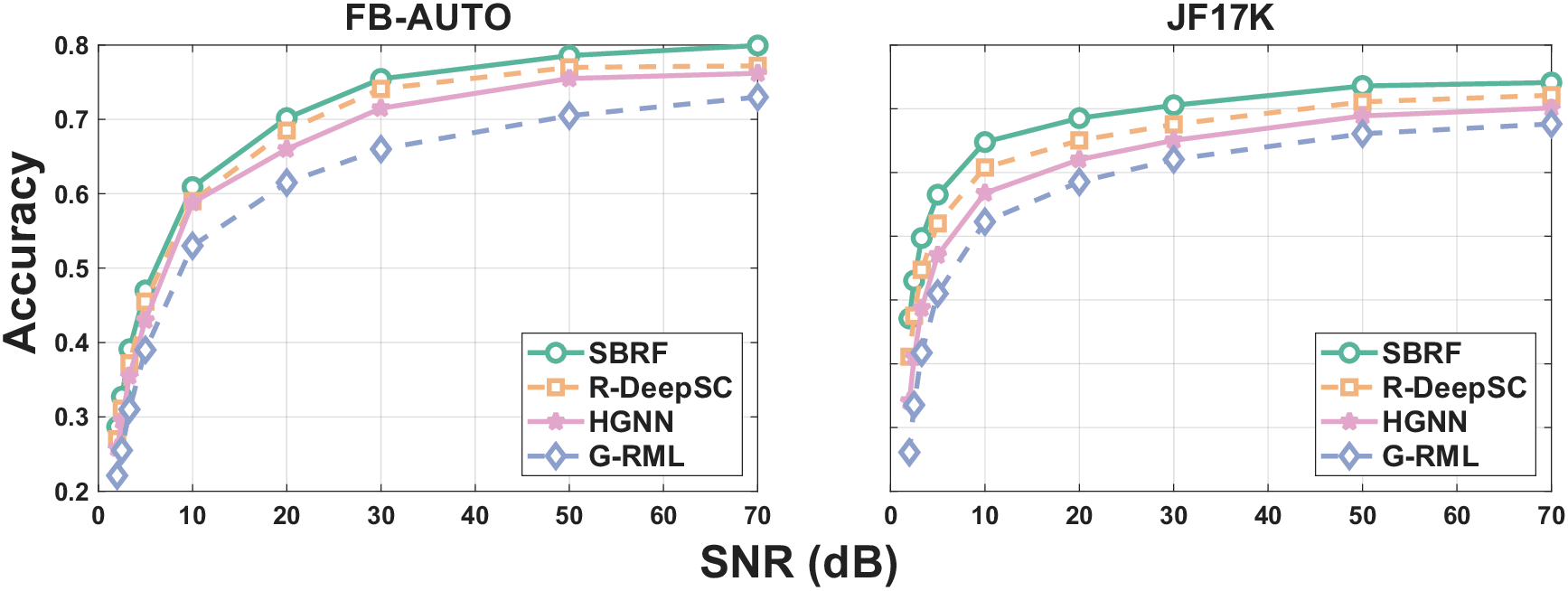}
    \caption{Performance comparison with baseline methods under different SNRs.}
    \label{fig:compare}
\end{figure}

In Fig.~\ref{fig:compare}, we compare the recovery accuracy of semantic information based on our proposed SBRF and three state-of-the-art semantic communication benchmarks, including transformer-based R-DeepSC\cite{peng2024robust}, joint-learning-based models HGNN\cite{gao2022hgnnplus}, and G-RML (TransE-based)\cite{xiao2023reasoning}, evaluated based on two datasets FB-AUTO and JF17K. We can observe that our proposed SBRF achieves the highest recovery accuracy among all the state-of-the-art solutions. Specifically, for dataset FB-AUTO, when the SNR is low, e.g., at 10 dB, our method outperforms R-DeepSC, HGNN, and G-RML by approximately 6.9\%, 8.3\%, and 17.5\%, respectively, in semantic recovery accuracy. Similarly, for dataset JF17K at the same SNR, our method achieves improvements of around 7.1\%, 9.4\%, and 19.8\% over R-DeepSC, HGNN, and G-RML, respectively. This further verifies the effectiveness of our proposed Bayesian posterior inference mechanism, which leverages structured prior knowledge to reconstruct the high-order implicit semantic information. 


\section{Conclusion}

This paper has proposed the SBRF, a novel Bayesian hypergraph inference-based semantic communication framework that can directly recover implicit semantic information involving high-order hyperedges at the receiver based on the pairwise relation-based explicit semantics sent by the transmitter. We have proved that the computational complexity of our proposed algorithm scales linearly with the maximum size of the recoverable hyperedge. Experimental results based on real-world datasets have shown that the proposed SBRF achieves up to 90\% recovery accuracy of the high-order hyperedges based on the pairwise relation-based explicit semantics.

\bibliographystyle{IEEEtran}
\bibliography{reference.bib}

\end{document}